\documentclass[12pt,epsf]{article}
\usepackage{graphicx}
\begin{document}

%%%%%%%%%%%%%%%%%%%%%%%%%%%%%%%%%%%%%%%%%%%%%%%%%%%%%%%%%%%%%%%%%%%%
%%%%%%%%%%%%%%%%%%%%%%%%%%%%%%  Defs. %%%%%%%%%%%%%%%%%%%%%%%%%%%%%%
%%%%%%%%%%%%%%%%%%%%%%%%%%%%%%%%%%%%%%%%%%%%%%%%%%%%%%%%%%%%%%%%%%%%

\def\a{\alpha}
\def\b{\beta}
\def\c{\varepsilon}
\def\d{\delta}
\def\e{\epsilon}
\def\f{\phi}
\def\g{\gamma}
\def\h{\theta}
\def\k{\kappa}
\def\l{\lambda}
\def\m{\mu}
\def\n{\nu}
\def\p{\psi}
\def\q{\partial}
\def\r{\rho}
\def\s{\sigma}
\def\t{\tau}
\def\u{\upsilon}
\def\v{\varphi}
\def\w{\omega}
\def\x{\xi}
\def\y{\eta}
\def\z{\zeta}
\def\D{\Delta}
\def\G{\Gamma}
\def\H{\Theta}
\def\L{\Lambda}
\def\F{\Phi}
\def\P{\Psi}
\def\S{\Sigma}

\def\o{\over}
\def\beq{\begin{eqnarray}}
\def\eeq{\end{eqnarray}}
\newcommand{\gsim}{ \mathop{}_{\textstyle \sim}^{\textstyle >} }
\newcommand{\lsim}{ \mathop{}_{\textstyle \sim}^{\textstyle <} }
\newcommand{\vev}[1]{ \left\langle {#1} \right\rangle }
\newcommand{\bra}[1]{ \langle {#1} | }
\newcommand{\ket}[1]{ | {#1} \rangle }
\newcommand{\EV}{ {\rm eV} }
\newcommand{\KEV}{ {\rm keV} }
\newcommand{\MEV}{ {\rm MeV} }
\newcommand{\GEV}{ {\rm GeV} }
\newcommand{\TEV}{ {\rm TeV} }
\def\diag{\mathop{\rm diag}\nolimits}
\def\Spin{\mathop{\rm Spin}}
\def\SO{\mathop{\rm SO}}
\def\O{\mathop{\rm O}}
\def\SU{\mathop{\rm SU}}
\def\U{\mathop{\rm U}}
\def\Sp{\mathop{\rm Sp}}
\def\SL{\mathop{\rm SL}}
\def\tr{\mathop{\rm tr}}

\def\IJMP{Int.~J.~Mod.~Phys. }
\def\MPL{Mod.~Phys.~Lett. }
\def\NP{Nucl.~Phys. }
\def\PL{Phys.~Lett. }
\def\PR{Phys.~Rev. }
\def\PRL{Phys.~Rev.~Lett. }
\def\PTP{Prog.~Theor.~Phys. }
\def\ZP{Z.~Phys. }

%%%%%%%%%%%%%%%%%%%%%%%%%%%%%%%%%%%%%%%%%%%%%%%%%%%%%%%%%%%%%%%%%%%%

\baselineskip 0.7cm

\begin{titlepage}

\begin{flushright}
%UT-
\end{flushright}

\vskip 1.35cm
\begin{center}
{\large \bf
511 keV Gamma Ray from Moduli Decay in the Galactic Bulge 
}
\vskip 1.2cm
M. Kawasaki${}^{1}$, and T. Yanagida${}^{2,3}$
\vskip 0.4cm

${}^1${\it Institute for Cosmic Ray Research, 
  University of Tokyo, Kashiwa, Chiba 277-8582, Japan}\\
${}^2${\it Department of Physics, University of Tokyo,
     Tokyo 113-0033, Japan}\\
${}^3${\it Research Center for the Early Universe, University of Tokyo,
     Tokyo 113-0033, Japan}

\vskip 1.5cm

\abstract{
We show that the $e^++e^-$ decay of a light scalar boson of 
mass $1-10$ MeV may 
account for the fluxes of 511 keV gamma ray observed 
by SPI/INTEGRAL.
We argue that candidates of such a light scalar boson 
is one of the string moduli or a
scalar partner of the axion in a supersymmetric theory. 

 }
\end{center}
\end{titlepage}

\setcounter{page}{2}

%%%%%%%%%%%%%%%%%%%%%%%%%%%%%%%%%%%%%%%%%%%%%%%%%%
\section{Introduction}
%%%%%%%%%%%%%%%%%%%%%%%%%%%%%%%%%%%%%%%%%%%%%%%%%%

It has been known in the superstring theory that the compactification 
with extra-dimensional fluxes stabilizes some moduli in supersymmetric 
(SUSY) string vacua \cite{Linde}. It is not yet, however, clear if
all of the moduli fields in the sting vacua are stabilized by the flux 
compactifiction. Therefore, it is a very important task to search  
a possible evidence for light scalar bosons. 
In this letter we point out 
that the 511 keV gamma-ray emission line from the galactic bulge
measured by the SPI spectrometer on the space observatory INTEGRAL
\cite{integral} is explained by the $e^++e^-$ decay of a
(pseudo)scalar boson of mass $O(1)$ MeV.\footnote{%%
The possibility of the 511keV gamma-ray emission due to
annihilation of dark matter particles was pointed out in Ref.~\cite{Boehm:2003bt}.}
We also argue that such a light
boson can be identified with one of the string moduli or 
with a scalar
partner of the Peccei-Quinn axion field in a SUSY theory. 
We also stress that the gauge-mediation model with a light 
gravitino of mass $O(1)$ MeV \cite{IY} is very interesting, 
since the above particles acquires most likely their masses of 
the order of the gravitino mass $m_{3/2}$ if they survive the 
flux compactification.

%%%%%%%%%%%%%%%%%%%%%%%%%%%%%%%%%%%%%%%%%%%%%%%%%%
\section{The $\phi\rightarrow e^++e^-$ decay}
%%%%%%%%%%%%%%%%%%%%%%%%%%%%%%%%%%%%%%%%%%%%%%%%%%

We consider a scalar boson $\phi$ which has a Yukawa coupling to the
electron as
\begin{equation}
   {\cal L}= \frac{m_e}{M_*}{\bar e}e\phi.
\end{equation}
If the $\phi$ is one of the string moduli, the $M_*$ is of 
the order of the
gravitational scale, $M_*\simeq M_G \simeq 2.4\times 10^{18}$ GeV
\cite{Witten}, and if it is a scalar partner (saxion) of the axion, 
$M_*\simeq F_a$ where $F_a$ represents the breaking scale of the
Peccei-Quinn symmetry. The decay width is given by
\begin{equation}
   \Gamma (\phi\rightarrow e^++e^-) \simeq \frac{1}{8\pi}
   \left(\frac {m_e}{M_*}\right)^2m_\phi,
\end{equation}
and we get the lifetime as
\begin{equation}   
   \label{eq:life}
   \tau (\phi\rightarrow e^++e^-) \simeq 2\times 10^{11} 
   \left(\frac{M_*}{10^{16}{\rm GeV}}\right)^2
   \left(\frac{{\rm MeV}}{m_\phi}\right)~ {\rm yr}.
\end{equation}
The lifetime of the $\phi$ boson should be longer than the age of the
universe ($\simeq 1.3\times 10^{10}$~yr) to give a significant flux 
of the gamma ray, which leads to a constraint 
\begin{equation}
   \label{eq:const}
   M_* \gsim (2.5-8.1)\times10^{15} {\rm GeV}~~~~~~~
   {\rm for}~ m_\phi \simeq 1-10~{\rm MeV}.
\end{equation}
Here we assume that the scalar $\phi$ does not have a direct
coupling to photons. The decay into neutrinos is 
negligible because of chirality suppression.  

%%%%%%%%%%%%%%%%%%%%%%%%%%%%%%%%%%%%%%%%%%%%%%%%%%
\section{The energy density of the $\phi$ boson}
%%%%%%%%%%%%%%%%%%%%%%%%%%%%%%%%%%%%%%%%%%%%%%%%%%

It is quite natural to consider that the $\phi$ boson has a large
classical value $\phi_0$ of the order $M_*$ at the end of inflation.
The $\phi$ starts a coherent oscillation when the Hubble constant 
of the universe reaches at the mass of the boson, $m_\phi$. 
The energy density of the coherent oscillation easily 
exceeds the  critical density of the 
universe~\cite{Coughlan:1983ci,Banks:1993en,deCarlos:1993jw}, which
leads to a serious cosmological problem (moduli problem). 
The thermal inflation \cite{Lyth:1995ka} is the most promising
mechanism to dilute the energy density
of the $\phi$ oscillation and 
hence solve the moduli problem. We
estimated in~\cite{Hashiba:1997rp,Asaka:1997rv} the energy density, 
$\Omega_\phi$, after the thermal inflation. 
The $\Omega_\phi$ in the present universe is given by
\begin{equation}
   \label{eq:omega}
   \Omega_\phi \gsim 4.0\times 10^{-4}
   \left(\frac{\kappa M_{*}}{M_{G}}\right)^2
   \left(\frac{m_{\phi}}{\rm MeV}\right)^{-3/14},
\end{equation}
for $m_{\phi} \lsim 10$~MeV. 
Here, we have taken $h\simeq 0.7$ ($h$: Hubble constant
in units of 100km/sec/Mpc) in eq. (20) of Ref.~\cite{Hashiba:1997rp} 
and 
assumed the initial amplitude of the $\phi$ 
to be $\phi_0=\kappa M_*$ with $\kappa$ being $O(1)$ constant.
The lowest density is realized for the reheating temperature 
after the thermal inflation
$T_R \simeq 10$~MeV (see Appendix and \cite{Hashiba:1997rp} 
for details).

%%%%%%%%%%%%%%%%%%%%%%%%%%%%%%%%%%%%%%%%%%%%%%%%%%
\section{The flux of the 511 keV gamma ray}
%%%%%%%%%%%%%%%%%%%%%%%%%%%%%%%%%%%%%%%%%%%%%%%%%%

Now we estimate the $511$~keV gamma ray flux from the Galactic center.
It was shown in Ref.~\cite{Hooper:2004qf} that $e^{+}+e^{-}$
decay of the dark matter particle of mass $m_{d}$ 
$\sim O(1-100)$~MeV can produce the 511~keV line emission  observed by 
SPI/INTEGRAL, through $e^{+}$ and (background) $e^{-}$ annihilation. 
Given a density $\Omega_{\phi}$, a mass $m_{\phi}$ and a lifetime
$\tau_{\phi}$ of the scalar particle,   
the $511$~keV gamma ray flux $\Phi_{511}$ is 
estimated as~\cite{Hooper:2004qf,Picciotto:2004rp}
\begin{equation}
   \Phi_{511} \sim 10^{-3} ~{\rm cm}^{-2}\sec^{-1}
   \Omega_{\phi}\left(\frac{10^{27}\sec}{\tau_\phi}\right)
   \left(\frac{\rm MeV}{m_\phi}\right),
\end{equation}
where we have used the present dark matter density 
$\Omega_{\rm dark} \sim 0.3$ and the halo density profile
$\rho_{\rm halo}\sim 1/r^{1.2}$\cite{Hooper:2004qf} 
($r$: distance from the Galactic center). 
Using  eqs.(\ref{eq:life}) and (\ref{eq:omega}) we obtain
\begin{equation}
   \label{eq:flux}
   \Phi_{511} \gsim 10^{-3} ~{\rm cm}^{-2}\sec^{-1}
   \kappa^2  \left(\frac{m_{\phi}}{\rm MeV}\right)^{-3/14}.
\end{equation}
It should be remarkable that the prediction of the 
flux is independent of $M_{*}$ and weakly depends on the 
mass of the scalar field.
Notice that the observed flux is~\cite{integral}
\begin{equation} 
   \Phi_{511} = 9.9^{+4.7}_{-2.1}\times 10^{-4}
   {\rm cm}^{-2}\sec^{-1}.
\end{equation}
We see, from 
eq.(\ref{eq:flux}), that  the positrons emitted by 
the decay of $\phi$ in the Galactic bulge explain 
naturally ($\kappa \simeq 1$) the observed fluxes of the 
511~keV line gamma ray if the thermal inflation maximally dilutes 
the scalar field density.

%%%%%%%%%%%%%%%%%%%%%%%%%%%%%%%%%%%%%%%%%%%%%%%%%%
\section{Conclusions}
%%%%%%%%%%%%%%%%%%%%%%%%%%%%%%%%%%%%%%%%%%%%%%%%%%

We have shown that the $e^{+}+e^{-}$ decay of a light scalar 
particle of mass $O(1)$~MeV 
diluted by the thermal inflation is capable of 
producing 511~keV gamma rays observed by the SPI/INTEGRAL. 

Since  the scalar field does not have a direct 
coupling to photons in the present model, the two photon decay 
($\phi\rightarrow 2\gamma$) takes place at most
through one-loop corrections and hence its branching ratio is
less than $\sim 10^{-6}$ for $m_\phi \sim 1$~MeV, 
which is well below the present observed
gamma ray background~\cite{Zdziarski}. 
However, the photon flux is only one or two oder of
magnitude smaller than the current limit for larger 
mass $\sim 10$~MeV.
The detection of line gamma rays with energy 
$m_{\phi}/2$ by future experiments will confirm the present model,
since the line gamma rays emitted by this process is distinctive 
of the model.
On the other hand,  Beacom {\it et al.}~\cite{Beacom:2004pe} 
pointed out that "internal bremsstrahlung" emission 
($\phi \rightarrow e^{+}e^{-}\gamma$)
exceeds the observed diffuse gamma ray flux from the Galactic center
unless $m_{\phi} \lsim 40$~MeV. 
This is one of the reason why we have assumed 
$m_\phi \simeq 1-10$~MeV.

We consider that a good candidate for such a light
scalar particle is one of the  string moduli for 
$M_* \simeq M_G$.
However, we also stress that the gamma ray flux is 
independent of $M_{*}$ as long as it satisfies 
eq.(\ref{eq:const}). 
Thus, the saxion is another candidate if the Peccei-Quinn scale 
$F_{PQ}=M_{*}$ is $\sim 10^{15}$~GeV.Notice that since the efficient thermal inflation requires the
reheating temperature as low as $O(1)$MeV the axion
density is also diluted. 
However, as shown in \cite{Kawasaki:1995vt} the axion with
$F_{PQ} \sim 10^{15}$~GeV explains the dark matter density of the universe
for the reheating temperature of $O(1)$ MeV.\footnote{%%
The present axion density is estimated as $\Omega_a \simeq 
\theta^2(F_a/10^{15}{\rm GeV})^2 (T_R/10{\rm MeV})$, 
where  $F_a$ is the Peccei-Quinn
scale and $F_a\theta$ with $\theta\sim O(1)$ is the initial 
amplitude of the coherent axion oscillation.}
Therefore, the axion may be the dark matter in the present universe. 

The above particles most likely acquire masses of the 
order of the gravitino mass $m_{3/2}$. The $O(1)$~MeV 
mass required in the present scenario is 
expected in the framework of gauge mediation models of 
SUSY breaking \cite{IY}.

As for baryon density of the universe,  the low 
reheating temperature makes baryogenesis very hard in general.
However, we consider that the late-time
Affleck-Dine mechanism~\cite{Affleck:1984fy} may 
work ~\cite{Stewart:1996ai,Jeong:2004hy}. \footnote{%%
The Affleck-Dine leptogenesis~\cite{Murayama:1993em} 
may also work if the $LH_u$ flat direction is used with 
a large cutoff scale. }
%%

%%%%%%%%%%%%%%%%%%%%%%%%%%%%%%%%%%%%%%%%%%%%%%%%%% 
\vspace{0.5cm}
\noindent
{\large\bf Appendix}
%%%%%%%%%%%%%%%%%%%%%%%%%%%%%%%%%%%%%%%%%%%%%%%%%%%

\noindent
In this appendix we estimate the moduli density after the entropy
production by the thermal inflation. The moduli density before 
the thermal inflation is written as 
\begin{equation}
  \frac{\rho_\phi}{s_i} ~\simeq ~\frac{m_\phi^2\phi_0^2}{T_\phi^3}
%  \simeq \frac{m_\phi^2\phi_0^2}{(m_\phi M_G)^{3/2}}
  ~\simeq ~ 10^{7.5}~{\rm GeV} 
  \left(\frac{m_\phi}{\rm MeV}\right)^{1/2}
  \left(\frac{\phi_0}{M_G}\right)^2,
\end{equation}
where $s_i$ is the entropy density before the thermal inflation
and $\phi_0$ and $T_\phi \simeq \sqrt{m_\phi M_G}$ is the 
amplitude and the cosmic temperature when the moduli field 
starts to oscillate. 

According to 
Refs.~\cite{Hashiba:1997rp,Asaka:1997rv}, we adopt the 
following potential of the flaton $\chi$ which is responsible 
for the thermal inflation:
\begin{equation}
   V(\chi) = V_0 + (T^2 -m_0^2)|\chi|^2 
   +\frac{\lambda^2}{M_G^2}|\chi|^6
\end{equation}
where $T$ is the temperature, $m_0$ is the order of the 
electroweak scale, $\lambda$ is a coupling constant, 
and $V_0 \simeq \lambda ^{-1}m_0^3 M_G$. 
For $T> m_0$ the potential takes minumum
at $\chi=0$. Moreover, if $T < V_0^{1/4}$, the vacuum energy
$V_0$ dominates over the radiation energy of the cosmic plasma.
Thus, for $m_0 < T < V_0^{1/4}$ an inflation (therml inflation)
takes place. The thermal inflation produces enormous entropy
with temperature $T_R$ 
after the flaton $\chi$ decay, and the ratio of the final entropy
$s_f$ to initial entropy $s_i$ is estimated as
\begin{equation}
   \Delta \equiv \frac{s_f}{s_i} = \frac{(4/3)V_0/T_R}{s_i}
   \simeq \frac{V_0}{T_R m_0^3} \simeq 
   \lambda ^{-1}\frac{M_G}{T_R}.
\end{equation}
Then, the moduli density is diluted by the entropy production
as
\begin{eqnarray}
   \Omega_\phi  & = & \left(\frac{\rho_\phi}{s_f}\right)
   \left(\frac{s_0}{\rho_{cr,0}}\right)
   = \left(\frac{\rho_\phi}{s_i}\right)
   \left(\frac{s_0}{\rho_{cr,0}}\right)
   \frac{1}{\Delta} \nonumber\\
   & \simeq & 4\times 10^{-4}\lambda
   \left(\frac{m_\phi}{\rm MeV}\right)^{1/2}
   \left(\frac{\phi_0}{M_G}\right)^{2}
   \left(\frac{T_R}{10{\rm MeV}}\right),
   \label{eq:BB}
\end{eqnarray}
where $\rho_{cr,0}$ and $s_0$ is the present critical and
entropy density. Notice that this value is almost the same as
the minimum moduli density given by eq.~(\ref{eq:omega}) for 
$m_\phi \simeq 1$~MeV.

However, the moduli dilution by the thermal inflation is
not so simple because the field value that gives the minimum
of the moduli potential is shifted by the Hubble induced mass term
during the thermal inflation. This shift produces a new oscillation
with amplitude $V_0 \phi_0/(m_\phi M_G)^2$, which leads to
\begin{equation}
   \label{eq:TH}
   \Omega_\phi  \simeq   10^{-6}\lambda ^{-1}
   \left(\frac{m_\phi}{\rm MeV}\right)^{-2}
   \left(\frac{\phi_0}{M_G}\right)^{2}
   \left(\frac{T_R}{10{\rm MeV}}\right)
   \left(\frac{m_0}{\rm GeV}\right)^{3}.
\end{equation}
Therefore, the total moduli density is determined from 
contributions of both eqs.~(\ref{eq:BB}) and (\ref{eq:TH}). 
Assuming that the falton decays into two gluons ($T_R 
\simeq 10^{-3}\lambda^{1/2} m_0$), we can obtain the minimum
moduli density (\ref{eq:omega}) by varing $m_0$ and $\lambda$~\cite{Hashiba:1997rp}.

\end{document}